\documentclass{ws-procs9x6}
\usepackage{amsmath,amssymb}

\newcommand{\ii}{\mathrm{i}}

\newcommand{\pd}{\partial}

\newcommand{\N}{\mathcal{N}}

\newcommand{\e}{\mathrm{e}}

\newcommand{\q}{\tilde{\theta}}
\newcommand{\sgn}{\mathop{\mathrm{sgn}}}

\begin{document}

\title{On the dynamics of BMN operators of finite size and the model
of string bits\footnote{\uppercase{T}his
work is supported by \uppercase{R}ussian grant for support of leading scientific
schools 2052.2003.1, NATO
Collaborative Linkage Grant PST.CLG. 97938,
INTAS-00-00254 grant,
RF Presidential grants MD-252.2003.02, NS-1252.2003.2, INTAS grant 03-51-6346,
RFBR-DFG grant 436 RYS 113/669/0-2, RFBR grant 03-02-16193 and the European Community's Human Potential
Programme under contract HPRN-CT-2000-00131 Quantum Spacetime.}}

\author{S.~Bellucci, 
\underline{C.~Sochichiu}\footnote{On leave from: \uppercase{B}ogoliubov
\uppercase{L}ab. \uppercase{T}heor. \uppercase{P}hys., \uppercase{JINR}, 141980
\uppercase{D}ubna, \uppercase{M}oscow \uppercase{R}eg., \uppercase{RUSSIA} and
\uppercase{I}nstitutul de \uppercase{F}izic\u a \uppercase{A}plicat\u a
\uppercase{A\c S}, str. \uppercase{A}cademiei, nr. 5, \uppercase{C}hi\c sin\u au,
\uppercase{MD}2028 \uppercase{MOLDOVA}.}}

\address{INFN -- Laboratori Nazionali di Frascati,\\
Via E. Fermi 40, 00044 Frascati, Italy
\\
E-mail: sochichi@lnf.infn.it}

\maketitle

\abstracts{We consider the discretization effects of a string-bit model simulating
the near-BMN operators in the super--Yang--Mills model. The fermionic sector of this model
is altered by the so called \emph{species doubling}. We analyze the possibilities to
cure this disease and propose an alternative formulation of the fermionic sector
free from the above drawbacks. Also we propose a formulation of string bits with exact
supersymmetry, which produces however an even number of continuous strings in the limit
$J\to\infty$.}

\section{Introduction}
AdS/CFT correspondence  is originally
formulated\cite{Maldacena:1998re,Gubser:1998bc} as a duality relation between
string theory on AdS$_5\times S^5$ space and conformal theory of $\N=4$ Super
Yang--Mills (SYM) on four dimensional Minkowski space, which is the boundary of
AdS (see Ref.\cite{Aharony:1999ti} for a review).

Being a true duality, AdS/CFT correspondence relates the weakly coupled SYM regime
to the strongly coupled string theory and \textit{vice versa}. This makes it into
a strong predictive tool, however a difficult ont to test, since there are no
means to probe either SYM or string theory at strong coupling. Beyond this, the
solution of string theory on AdS background is not known (although some progress
towards this was recently achieved, see Ref.\cite{Bena:2003wd} and references
therein).

A few years ago, Berenstein--Maldacena--Nastase%
\cite{Berenstein:2002jq,Berenstein:2002sa,Berenstein:2002zw} proposed to
consider the particular limit of AdS$_5\times S^5$ geometry known as pp-wave
background, on which the string theory is solvable%
\cite{Metsaev:2001bj,Metsaev:2002re}. This situation of the overlap of the
applicability of both string theory and SYM was further extended to spinning
string backgrounds\cite{Frolov:2003qc,Arutyunov:2003uj,Tseytlin:2003ii} (see also
the contribution of G. Arutyunov to the present proceedings).

In the absence of a more adequate tool to describe string theory near the BMN
limit it was proposed to use the string bit model%
\cite{Verlinde:2002ig,Vaman:2002ka,Zhou:2002mi}. This model consists of the set of
point-like interacting particles --- string bits which in the limit of the number
of particles going to infinity is expected to produce the continuous pp-wave
string. Later on, however, the exact description in terms of spin chains was found
in the planar limit of SYM \cite{Minahan:2002ve,Beisert:2003yb}. Nevertheless, string
bits remained an efficient tool to take care of nonplanarity which correspond, via
AdS/CFT, to string production.

Although the string bit model was quite useful in describing the dynamics of the bosonic
sector in the near BMN limit it suffers from internal inconsistencies due to the
fermionic spectrum doubling. In the theory of discrete fermions there is a
well-known no-go theorem due to Nielsen--Ninomiya \cite{Nielsen:1981hk} which
limits the possibilities of having discrete fermions. Therefore,  one should make
sure that these limits do not interfere with the AdS/CFT correspondence. For this purpose, one
needs to prove the existence of a discrete supersymmetric model which in the
continuum model reduces to the pp-wave string. This is the aim of the present note. It is
based mainly on the refs \cite{Bellucci:2003hq,Bellucci:2003qi} where this problem
was approached (for a staggered fermion approach see also \cite{Danielsson:2003yc}).

\section{Fermionic doubling}

After fixing the permutation symmetry, the one-string sector of the string bit
model is given by the following Hamiltonian:
\begin{multline}\label{hamiltonian}
  H_{sb}=\sum_{n=0}^{J-1}\left[\frac{a}{2}(p_{in}^2+x_{in}^2)
  +\frac{1}{2a}(x^i_{n+1}-x^i_n)^2\right]\\
  -\frac{\ii}{2} \sum_{n=0}^{J-1}\left[(\theta_n\theta_{n+1}-
  \q_n\q_{n+1})-2a\q_n\Pi\theta_n\right],
\end{multline}
with commutation relations
\begin{equation}\label{PB}
  [p^i_n,x^i_n]=-\frac{\ii}{a}\delta^{ij}\delta_{mn},\quad
  \{\theta^a_n,\theta^b_m\}=\frac{1}{a}\delta^{ab}\delta_{mn},\quad
  \{\q^a_n,\q^b_m\}=\frac{1}{a}\delta^{ab}\delta_{mn},
\end{equation}
where $i=1,\dots,8$ are vector and, respectively, $a=1,\dots,8$ spinor indices of
$SO(8)$ appearing in the light-cone quantization of the pp-wave string. $\Pi$ is
the matrix in $SO(8)$ spinor space given in terms of $16\times 16$ dimensional
$\gamma$-matrices in chiral representation by
\begin{equation}\label{Pi}
    \Pi=\gamma^1\bar{\gamma}^2\gamma^3\bar{\gamma}^4,\qquad \Pi^2=1.
\end{equation}

The Hamiltonian \eqref{hamiltonian}, together with the shift operator
\begin{equation}\label{p}
  P=\frac{a}{2}\sum_n \left[2p^i_{n}\pd x_{n}+\ii
  (\theta_{n}\pd\theta_{n}+\q_{n}\pd \q_{n})\right],
\end{equation}
where $\pd f_n=(1/a)(f_{n+1}-f_n)$ are generated by the ``supercharges''
\begin{equation}\label{susy}
    \{Q_a,Q_b\}=2\delta_{ab}(H+P), \quad
    \{\tilde{Q}_a,\tilde{Q}_b\}=2\delta_{ab}(H-P),
\end{equation}
where
\begin{subequations}\label{q-ansatz}
  \begin{align}
    Q&=a\sum_n\left[p^i_{n}\gamma^i\theta_{n}-
    x^i_{n}\gamma^i\Pi\q_{n}+\pd x_{n}\gamma^i\theta_{n}\right],\\
    \tilde{Q}&=a\sum_n\left[p^i_{n}\gamma^i\q_{n}+
    x^i_{n}\gamma^i\Pi\theta_{n}-\pd x_{n}\gamma^i\q_{n}\right].
  \end{align}
\end{subequations}
In spite of the fact that the supercharges generate according to \eqref{susy} both the Hamiltonian
and the shift operator, the supersymmetry fails because the supercharges do not commute
\begin{equation}
  \{Q,\tilde{Q}\}\neq 0.
\end{equation}

The fermionic doubling is better seen if we pass to the Fourier mode description
\begin{equation}
  f_{n}=\frac{1}{\sqrt{J}}\sum_{k=-J/2}^{J/2-1}f_k\e^{2\pi\ii
  kn/J},
\end{equation}
where $f_n$ stands for $x_n$, $p_n$ $\theta_n$ and $\q_n$ while $f_k$ represents
their Fourier modes. In terms of Fourier modes the fermionic part of the
Hamiltonian has the form
\begin{equation}\label{H_f}
    H_f=\frac{1}{2}\sum_k\left[
    \sin\frac{2\pi k}{J}(\theta_{-k}\theta_{k}-\q_{-k}\q_{k})-2a\ii
    \q_{-k}\Pi\theta_k\right],
\end{equation}
where the doubling is visible due to additional zeroes of the sin function in eq.
\eqref{H_f}. Due to the fact that modes around these additional zeroes have low
energy, they survive and
contribute in the continuum limit, though they correspond to fast oscillating fields.
In contrast, in the bosonic part one has a
factor with the cosine function which becomes large in the continuum limit and one has
no doubling there. Therefore, contrary to what can be expected, supersymmetry is not
restored for $a\to 0$.

In \cite{Danielsson:2003yc} it was proposed to use a staggered fermion approach in
order to cure the fermion doubling. It consists in placing only a half number of
fermions, e.g. in odd points ($n=2k+1$) one has only $\theta_n$  while in even
($n=2k$) there are only $\q_n$. Due to doubling the number of fermions in the
continuum limit is the correct one. The supersymmetry is also restored in the
continuum limit. However, at any finite lattice size it is still broken.

\section{Almost supersymmetry}

In fact, it appears possible to make the theory supersymmetric with the exception
of a single mode by properly choosing the lattice derivative function. If this
mode is placed in the high energy region of the model one may expect that it does
not contribute in the continuum limit.

On the lattice one can have various definitions of the derivative of a field. The
only criterion is that the lattice derivative should give the usual derivative in
the continuum limit. In general the lattice derivative could be written as the operator
\begin{equation}\label{lat-der}
  \bar{\pd}f_n=\sum_{m}\bar{\pd}_{nm} f_m,
\end{equation}
where $\bar{\pd}_{nm}\equiv\bar{\pd}_{n-m}$ are the elements of the derivative matrix
$\bar{\pd}$. The condition that $\bar{\pd}_{n}$ describes a lattice derivative reads
\begin{equation}
  \sum_{m}\bar{\pd}_{m}=0,\qquad \sum_{m}\bar{\pd}_{m}m=1,
\end{equation}
and
\begin{equation}
  \lim_{a\to 0}\frac{a^k}{k!}\sum_{m}\bar{\pd}_{m}m^k\to 0,\qquad k\geq 2.
\end{equation}
In terms of the Fourier mode expansion this means that the Fourier transform
$\hat{\pd}_{k}$ should have the limit
\[
  \bar{\pd}_{k}|_{a\to 0}\to \ii p_k=2\pi\ii k.
\]

Substituting the forward lattice derivatives $\pd$ in \eqref{q-ansatz} by a
generic derivative $\bar{\pd}$ and requiring that we get the same bosonic part of
the Hamiltonian, we find the condition of restoration of supersymmetry to be
\begin{equation}\label{root}
    \bar{\pd}=\ii(\pd^*\pd)^{\frac{1}{2}}.
\end{equation}

The condition \eqref{root} is trivially satisfied in the continuum limit, since (in Fourier
mode) $\pd_k= -\ii p_k$ is an anti-Hermitian operator. In contrast, on the lattice
$\pd$ has both Hermitian and anti-Hermitian parts.\footnote{In general, the
Hermitian part of the derivative is the one responsible for fermionic doubling.}
Therefore, extracting the square root in \eqref{root} is a nontrivial procedure also
because of its ambiguity.

In terms of Fourier modes equation \eqref{root} is reduced to the extraction of the
square root for each mode
\begin{equation}\label{signs}
    \bar{\pd}_{k}=\ii\epsilon_k\sqrt{\frac{1}{2}\left(\cos \frac{2\pi
    k}{J}-1\right)}=\ii\epsilon_k\left|\sin \left(\frac{\pi k}{J}\right)\right|,
\end{equation}
where $\epsilon_k=\pm 1$ is the sign, which can be chosen separately for each mode.
The natural choice $\epsilon_k=\sgn k$ yields a jump in the derivative function at $k=\pm
J/2$. Also at this point the solution \eqref{signs} fails to satisfy equation
\eqref{root}. The jump of the derivative function is related to the \emph{nonlocality} of
the fermionic derivative
\begin{equation}\label{x-reps}
  \bar{\pd}_{nm}=\frac{2}{a}(-1)^{n-m}\frac{
    \cos (\frac{\pi }{2J})
    \sin (\frac{\pi(n-m) }{J})}
    {\cos (\frac{\pi }{J}) -
    \cos (\frac{2\pi (n-m)}{J})}.
\end{equation}
Although \eqref{x-reps} looks nonlocal and ugly, one can check\footnote{The
simplest way to do this is to use the Fourier transform.} that its square is a
local operator given by
\begin{equation}\label{square}
  (\bar{\pd}^2 f)_n=(f_{n+1}-2f_n+f_{n-1})/a^2,
\end{equation}
which is the next-to-neighbor second derivative.

Let us see that the  ambiguity of the momentum function does not affect physical
observables like the energy.

\section{Fermion doubling/nondoubling in the BMN correspondence}

Earlier we argued that having a well-defined supersymmetric discrete model is
indispensable for confirming the self-consistence of the BMN limit in the fermionic
sector. Now we want to define precisely the quantities to be compared.

In the mode expansion of the pp-wave string described by Metsaev
\cite{Metsaev:2001bj,Metsaev:2002re} one has the energy of each mode given by the
square root
\begin{equation}\label{en_root}
  \omega_n\sim \sqrt{1+\lambda^2n^2},
\end{equation}
where $\lambda$ is the 't Hooft coupling (for simplicity, other parameters here
are put to unity). On the other hand, the Yang--Mills contribution at $k$ loops
yields a quantity of order $\sim \lambda^{2k}$ which corresponds to the $k$-th
term in the expansion of the square root \eqref{en_root}. The planar Yang--Mills
contribution up to three loops nowadays is well known and is described by
integrable spin chains\cite{Minahan:2002ve,Beisert:2003yb,Beisert:2003ys}.

In the string bit model the argument of the square root is replaced by
\cite{Bellucci:2003hq,Bellucci:2003qi}
\begin{equation}
  \bar{\omega}_n\sim \sqrt{1+\lambda^2\bar{\pd_k}^2}.
\end{equation}

Let us consider, for definiteness, the contribution of order $\lambda^2$
which corresponds to the Yang--Mills one-loop approximation. Then the energy of the
string bit mode in this approximation is given by
\begin{equation}\label{1-loop-en}
  \omega^{1-loop}_k\sim \lambda^2\bar{\pd}_k^2=\lambda^2(\pd^*\pd)_k,
\end{equation}
where $\pd$ is the usual bosonic forward lattice derivative $\pd
f_n=(1/a)(f_{n+1}-f_n)$, while $\pd^*$ is the backward one
$\pd^*f_n=(1/a)(f_n-f_{n-1})$.

The one loop energy \eqref{1-loop-en} corresponds to the following
term in the fermionic Hamiltonian:
\begin{multline}\label{dpsi_quad}
  \sim
  -a \sum_n\psi^{(+)}_n\bar{\pd}^2\psi^{(-)}_n\\
  =a\sum_n\pd\psi^{(+)}_n\pd\psi^{(-)}_n
  =a^{-1}\sum_n(\psi^{(+)}_{n+1}-\psi^{(+)}_{n})(\psi^{(-)}_{n+1}-\psi^{(-)}_{n}),
\end{multline}
where we integrated by parts after using \eqref{square}, in order to express
$\bar{\pd}^2$ in terms of next neighbor derivatives $\pd$ and
$\pd^*$.

Let us note that the fields $\psi^{(\pm)}_n$ entering in \eqref{dpsi_quad} are
inverse Fourier transforms of Hamiltonian eigenmodes rather than the original
fermionic fields $\theta_n$ and $\tilde{\theta}_n$. The transformation which relates
them is a singular one, what makes it possible for the Hamiltonian to be ill-defined
in terms of $\theta_n$ and $\tilde{\theta}_n$, although in terms of $\psi^{(\pm)}_n$
there is no such problem. In particular, this means that the eigenvalues of the
fermionic Hamiltonian are well-defined and compatible with supersymmetry in spite
of the discontinuity in \eqref{signs}.

\section{Doubling and supersymmetry}

In the last section we have shown that one can consistently define the supersymmetric
model with the right spectrum, provided one does not insist on the worldsheet fermionic
structure. Let us show now that one can have more. Namely, let us construct an
exactly supersymmetric string bit model.

As we have established, supersymmetry is present when the fermionic discrete
derivative $\bar{\pd}$ and the bosonic one $\pd$ satisfy the relation
\eqref{root}. The problem arises when we try to solve equation \eqref{root} for the
fermionic $\bar{\pd}$ with a given bosonic $\pd$, corresponding to the next
neighbor lattice derivative. In this section we want to show that the problem
disappears if one relaxes the latter condition and allows also for an arbitrary bosonic
derivative with unambiguous square root \eqref{root}. Since such a derivative
should vanish at the end of the Brillouin zone ($k=\pm J/2$, in the case of an even
number of bits), the bosonic sector should be also doubled! In other words, since
we are unable to avoid the fermionic doubling without the alteration of the model
let us allow the doubling also for the bosons, then one can expect to have
supersymmetry.

The simplest choice for a ``fermionizable'' bosonic derivative function $\pd_k$ can
be obtained by choosing it to be the ``doubled'' fermionic one
\begin{equation}\label{new-bosonic-d}
  \pd^{new}_k=\frac{\ii}{2a}\sin\left(2\pi k a\right).
\end{equation}
This corresponds to the symmetric derivative of the bosonic fields,
\[
  \pd^{new}f_n=(1/2a)(f_{n+1}-f_{n-1}).
\]
Since the derivative \eqref{new-bosonic-d} is purely imaginary, it solves also
eq. \eqref{root} for the fermionic derivative: $\bar{\pd}=\pd^{new}$. Then, the
supersymmetric Hamiltonian is a combination of the bosonic part with derivative
\eqref{new-bosonic-d} and the naive fermionic part
\begin{multline}\label{new-hamiltonian}
  H_{new}=\sum_{n=0}^{J-1}\left[\frac{a}{2}(p_{in}^2+x_{in}^2)
  +\frac{1}{8a}(x^i_{n+1}-x^i_{n-1})^2\right]\\
  -\frac{\ii}{2} \sum_{n=0}^{J-1}\left[(\theta_n\theta_{n+1}-
  \q_n\q_{n+1})-2a\q_n\Pi\theta_n\right],
\end{multline}
which generates the supersymmetry algebra \eqref{susy} together with the
supercharges \eqref{q-ansatz} and the shift operator \eqref{p}, where we put
$\pd=\bar{\pd}=\pd^{new}$.

Let us note that no even bit $n$ in \eqref{new-hamiltonian} interacts with an odd
one in both fermionic and bosonic sectors. Therefore, the model is doubled in the bosonic
sector as well as in the fermionic one. This is confirmed by the inspection of the
bosonic Hamiltonian which now reads
\begin{equation}\label{new-bos-ham}
  H^{new}_{bosonic}=\sum_{k}\left[\frac{1}{2}|p_k|^2+\left(1+\frac{1}{8}\sin^2(2\pi
  ka)\right)|x_k|^2\right].
\end{equation}

As it can be seen from Eq. \eqref{new-bos-ham}, $\sin^2(2\pi ka)$ has two zeroes~:
one at the origin $k=0$ and one at the edge of the Brillouin zone $k=\pm J$. Each
of these zeroes contributes in the continuum limit a superstring. Therefore one
ends up with two superstrings.

One can consider a derivative with any even number $2s$ of zeroes of the derivative
function $\pd_k$. In this case one will have $2s$ superstrings in the continuum
limit. Let us note that, since the supersymmetry is present at any stage of the
discrete model, the continuum theory is also supersymmetric.

\section{Discussion}
In this note we reviewed the fermionic spectrum of the string bit model. In the naive
formulation the model suffers from inconsistencies due to the fermion spectrum
doubling and supersymmetry breaking. Supersymmetry is not restored
automatically in the continuum limit.

One can ``optimize'' the fermionic sector for the given form of the bosonic
Hamiltonian, in order for supersymmetry breaking to be minimal. In fact, one can
have supersymmetry well-defined on all modes but one placed on the edge of the
Brillouin zone. We have shown that this discrepancy does not affect, however, the
physical spectrum of the model, since it depends on the square of the fermionic
derivative which is a well-defined function.

In the case one allows doubling also for bosonic part, one can construct a fully
supersymmetric discrete theory. The continuum limit of this model is expected to
contain a pair or, in general, any even number of superstrings. It is interesting
to find out if one can use a correlated doubling removing procedure for both the
bosonic and the fermionic sector, like the Wilson generalization or some other procedure, in
order to get a single copy of string in the continuum limit and preserve
supersymmetry. We hope to return to this elsewhere.

Finally, our analysis leads us to the conclusion that the obstructions on the existence
of discrete supersymmetric models are not enough to make the BMN limit in
SYM theory problematic. Beyond this, there is no topological nor any
other obstruction to this, since the theory is not chiral. In this sense, this case
is similar to $SU(2)$ chiral fermions on the lattice, where one can have a discrete
theory preserving all perturbative symmetries\cite{Sochichiu:1998bg}.

\section*{Acknowledgement}
We thank G.~Arutyunov, N.~Beisert and J.~Plefka for useful discussions and the
organizers of the Conference for warm hospitality and creative atmosphere.

\providecommand{\href}[2]{#2}\begingroup\raggedright\endgroup

\end{document}